\documentclass{svproc}
\usepackage{url}

\usepackage{amsmath, epsfig}
\usepackage{amssymb}
\usepackage{mathrsfs}
\usepackage{hyperref}
\usepackage[latin1]{inputenc}
\usepackage{color}

\usepackage{cancel}
\usepackage{soul}
\usepackage{ulem}
\usepackage{amsfonts}
\usepackage{amstext}
\usepackage{graphicx}
\usepackage{amscd}
\usepackage{amsmath}
\usepackage{mathrsfs}
\usepackage{enumerate}
\usepackage{cite}

\usepackage[font=footnotesize]{caption}
\usepackage{floatrow}
\usepackage{subfigure}
\usepackage{sidecap}
\usepackage{graphicx}

\newcommand{\llangle}{\langle \hspace{-0.2em} \langle}
\newcommand{\rrangle}{ \rangle \hspace{-0.2em}  \rangle}
\newcommand{\transp}{\text{\textsc{t}}}
\newcommand{\tr}{\mathrm{Tr}}

\newcommand{\Ket}[1]{\vert #1 \rrangle}

\begin{document}

\mainmatter              
\title{Transient synchronization in open quantum systems}
\titlerunning{Transient synchronization in open quantum systems}  
%
\author{Gian Luca Giorgi \and Albert Cabot \and Roberta Zambrini}

\authorrunning{G. L. Giorgi  et al.} 
\institute{IFISC (UIB-CSIC), Instituto de Fisica Interdisciplinar y Sistemas Complejos Universitat de les Illes Balears-Consejo Superior de Investigaciones Cientificas, 
UIB Campus, E-07122 Palma de Mallorca, Spain}

%
%
%

\maketitle              

\begin{abstract}

The phenomenon of spontaneous synchronization arises in a broad range of  systems when the mutual interaction strength  among components
overcomes the effect of detuning.
Recently it has been studied also in the quantum regime with a variety of approaches and in different dynamical contexts.
We review here transient synchronization arising during the relaxation of open quantum systems, describing the common enabling mechanism 
in presence of either local or global dissipation. We address both networks of harmonic oscillators and spins and compare different synchronization measures. 

\end{abstract}

\section{Introduction}
Spontaneous synchronization (SS) can be defined as the phenomenon where two or more individual systems mutually interact with each other so as to adjust their own local 
dynamics to a common pace, due to their interaction. It differs from the so-called entrainment, where a system resonates with an external periodic signal and follows its pace. 
The study about synchronization started in 1657 when Christian Huygens realized the
first pendulum clock. Among other fields, SS is commonly observed in physics, in engineering, in  biological and chemical systems, as well as in 
social sciences \cite{Pikovsky2001,Strogatz2001,Arenas,Manrubia,Rohden}.


The study of SS in the quantum regime has become a major research subject in the last decade in systems ranging from opto-mechanical oscillators
to atoms, in self-sustained as well as in relaxation dynamics, from bosons to spins, both in continuous and discrete variables (see review 
\cite{ZambriniRev} and references therein). 
In particular, the same definition and quantification of SS allows for different approaches in the quantum regime. 
As a matter of fact, 
synchronization in classical systems is a property of temporal trajectories. Lacking an immediate analogue of trajectories for coupled systems in the quantum regime,
it is not surprising the diversity of definitions of quantum SS. 
Different measures of quantum SS heve been  considered and compared with other forms of correlations in Ref. \cite{ZambriniRev}. 
The two main approaches concern either the study of time correlation of the dynamics of local observables  or the reduction of noise in  collective variables, 
which represent by definition a kind of global quantum correlations. 
These measures of synchronization provide a generalization of the classical ones and allow to explore  whether this phenomenon displays purely quantum effects.

Another interesting question is about the possibility of observing SS in different dynamical regimes.
A largely explored framework, both in classical and quantum systems, deals with self-sustained oscillators \cite{Pikovsky2001,ZambriniRev} . These are autonomous systems in which an oscillatory dynamics emerges in the presence of non-linearity and a balance between input and output energy provided by driving and dissipation.
Synchronization is well established also beyond regular dynamics, as in chaotic systems, that can evolve identically in spite of their characteristic sensitivity to differences in initial conditions \cite{Strogatz2001}. This has been shown to
allow, for instance, for cryptography and communications based on chaotic signals \cite{Argyris}. 
Furthermore SS can arise also in the transient evolution of dynamical systems, during  relaxation towards equilibrium 
\cite{LeHur,SyncHO,Benedetti,manzanoPRA,GLG1,manzanoSciRep,GLG2,Bellomo,Biel,Cabotnetw,Olaya-Castro,Solano1,Solano2,militello1,militello2,Karpat,Cabot_atoms}.  
In particular  SS is recognized as a universal phenomenon in non-linear sciences \cite{Pikovsky2001} but it can also occur in linear systems \cite{SyncHO,manzanoPRA,manzanoSciRep}.

Here,  we aim to review some recent results on transient synchronization in quantum systems among harmonic oscillators and spins in contact with different kinds of environments and evolving towards a stationary steady state.
The  emergence of SS is induced by a clear and general mechanism of separation of multiple dissipative time scales in the dynamics \cite{SyncHO,GLG1,manzanoSciRep}. If the decay rates of the normal modes of 
the systems are such that one of such modes (or else a bunch of 
frequency degenerate modes) is much slower than the other ones, the predominant contribution to the long-time dynamics is represented by such decaying mode leading to 
a synchronous dynamics. 

Dissipation and noise play a key role in enabling this transient synchronization in open (quantum) systems, and we will review under which conditions this occurs, also discussing when local and global dissipation can induce SS. Beyond the results reviewed here,  dissipative couplings induced by a common bath of phonons  can enable SS in self-sustained oscillators even in the absence of reactive couplings among opto-mechanical systems \cite{cabotOM}.
 The possibility of having noise playing a constructive role for synchronization phenomena has been reported also in different contexts,
 for instance in noise-induced synchronization, and has been debated in classical chaotic systems  \cite{Noise-induced,toral},
 when starting from different initial conditions but considering a common noise source. 
Another context in which noise plays a constructive key role is stochastic resonance, where noise allows for entrainment to an external driving force  \cite{gammaitoni}. This phenomenon has also been generalized into the quantum regime   \cite{Grifoni}. 

With respect to SS in autonomous systems, transient synchronization in out-of-equilibrium systems has been by far less explored \cite{LeHur}.
Collective dissipation allows one to synchronize detuned oscillators 
\cite{SyncHO,Benedetti,manzanoPRA,manzanoSciRep,Cabotnetw}
and spins \cite{GLG1,GLG2,Bellomo}. In fact transient SS can be related to the phenomenon of subradiance
\cite{Bellomo}, while collective dephasing does not enable SS
\cite{GLG1}. On the other hand, also dissipation acting locally enables synchronization under proper conditions, as shown in networks
\cite{manzanoSciRep,militello1} and atomic lattices \cite{Cabot_atoms}.
The emergence of transient environment-induced SS between a pair of qubits has been recently explored in the framework of a collision model,  allowing for interactions and delays  among separate environments of spins  \cite{Karpat} .

A first example of an application based on transient SS has been proposed in the context of quantum probing: the transition between synchronization in phase and in antiphase can allow one  to probe the environment of a dissipating qubit with an external probe \cite{GLG2}.
This has also been exploited to improve the performance of probing assisted by machine learning \cite{Biel}.
When increasing the complexity of the system, as in random or small-world networks, transient SS is also found to be a persistent phenomenon \cite{Cabotnetw} that can be  triggered by local parameter tuning \cite{manzanoSciRep}.
Transient SS has been explored in relation to coherence in biomolecules considering a vibronic dimer in Ref. \cite{Olaya-Castro}. Hybrid spin-boson structures have also been considered in Ref. \cite{militello2} showing SS of oscillators mediated by dissipative spins and in Ref. \cite{Solano1}, also proposing to quantify the degree of quantumness in SS, through assessment of non-commuting observables.

In Sec. \ref{secii}, we discuss in general the problem of transient synchronization for both spin networks and harmonic oscillator networks, establishing the criteria for SS to be observed. In Sec. \ref{seciii}, we will review the measure of SS based on time correlation of local trajectories. In Sec. \ref{seciv}, we will explicitly discuss different examples of SS, for each of such examples pointing out an important feature: starting with the case of two harmonic oscillators, we will analyze the differences arising from dissipating in either a common bath or in separate environments; then, introducing a larger network we will discuss the possibility of transforming SS from a transient one to a  stationary because of the presence of noiseless channels; then,  we will first discuss the possibility of using either a local or a global approach to obtain the master equation in coupled spin systems (showing that when a local approach is possible SS cannot emerge), and then the role of pure dephasing. Finally, in Sec. \ref{secv} we will conclude this chapter.

\section{Models for transient synchronization}\label{secii}

Let us consider a generic $n$-partite system obeying a Hamiltonian ($\hbar=1$ in all the text)
\begin{equation}\label{eq1}
H_S=\sum_{i=1}^n \omega_i H_i+\sum_{i,j}\lambda_{ij}V_{ij},
\end{equation}
where $H_i$ are local diagonal Hamiltonians and $V_{ij}$ are coupling terms. The system is
interacting with an environment such that the reduced density matrix of the system $\rho$ obeys a Lindblad  master equation \cite{bp}
\begin{equation}\label{MasterEquation}
\dot{\rho}=-i [H_S,\rho]+\sum_\mu [2 L_\mu \rho L_\mu^\dag- \rho L_\mu^\dag L_\mu- L_\mu^\dag L_\mu \rho],
\end{equation}
where the operators $L_\mu$ can be either local or global depending on the kind 
of system-bath interaction. The Hamiltonian (\ref{eq1}) can describe either 
discrete-variable systems, as, for instance $\frac{1}{2}$-spins (qubits), or 
harmonic oscillators. 
In the absence of any driving force, such a dissipative system is expected to 
decay and end up in its steady state. 

\subsection{Global versus local description of the master equation}

As SS emerges among coupled (either directly or indirectly) parties interacting with environments that can in turn act either locally or globally, an important issue arises about the way of deriving the master equation in  different scenarios. 
For the sake of concreteness, let us consider the case of two spins interacting through separate environments described by the system Hamiltonian 
\begin{equation}
    H_S=\frac{\omega_0}{2}(\sigma_1^z+\sigma_2^z)+\lambda (\sigma_1^+\sigma_2^-+\sigma_2^+\sigma_1^-)
\end{equation}
and the interaction Hamiltonian
\begin{equation}
  H_I=\sigma_1^x B_1+\sigma_2^x B_2
\end{equation}
where $B_i$ are bath operators. Intuitively, in the limit of $\lambda$ smaller than the dissipative time scale fixed by the system-bath interaction, the perturbative Born-Markov master equation can be derived considering the interaction picture with respect to $\frac{\omega_0}{2}(\sigma_1^z+\sigma_2^z)$ \cite{cresser,scala,scala2,migliore,heinrich,wichterich,Trushechkin,correa,brunner}. In this limit, at zero temperature and under secular approximation,  the master equation would simply read
\begin{equation}\label{MasterEquationlocal}
\dot{\rho}=-i [H_S,\rho]+\sum_{i=1}^{2} {\cal D}_i[\rho]
\end{equation}
where the dissipative superoperator
\begin{equation}
{\cal D}_i[\rho]=2\gamma_i[\sigma_i^- \rho \sigma_i^+-\{\rho,\sigma_i^+ \sigma_i^-\}]
\end{equation}
accounts for the dissipation of each spin with rate $\gamma_i$,
independent of $\lambda$ (a microscopic derivation can be found in Refs.\cite{bp,marco}). Here it is important to remark that the local character of the master equation arises in the limit of small $\lambda$ and it is not necessarily implied by  the assumption of separate baths \cite{correa,brunner,marco}.
Indeed, in the cases where $\lambda$ cannot be treated as a perturbation, a proper derivation of the master equation leads to global jump operators. An explicit example will be discussed in Subsec. \ref{spinpairs}.

As a final remark, it is worth stressing that, in general, a global approach is necessary also in the case of a single local environment, as for instance, in Ref. \cite{GLG2}. Also, while in the limit of small inter-coupling the local approach of Eq. (\ref{MasterEquationlocal}) will give a valid dynamical description,
there is no regime where the global approach is less accurate  than the local one, 
provided that the secular approximation is carried out consistently. Indeed, the formal derivation of the master equation in general needs to  take into account the full system Hamiltonian and this leads to a global master equation, that under proper (partial) secular approximation describes accurately the dissipation of coupled spins.
This point is extensively discussed elsewhere \cite{marco}.



\subsection{Enabling mechanism: time scale separation}

 The specific form of dissipation can introduce a time scale separation in the modes of the composed system that can provide 
a rather general mechanism for mutual synchronization. 
If, during the dynamical relaxation, one of the normal modes of the system decays much slower than 
any other mode, there will be a large transient in which
only oscillations at the frequency of the slower mode will be observed looking 
at the dynamics of any local subsystem. 
Then, such subsystems will result to be dynamically synchronized. Obviously, 
such a  mechanism relies on the existence of dissipation, which is the physical 
process 
that filters out all the modes but one, while it has been shown that pure dephasing does not 
allow for SS \cite{GLG1}. A further fundamental ingredient is 
represented by the existence of a gap in the  eigenmode decay rates between the   
slowest one and the second to last, when they correspond to different frequencies. This allows for the existence of a significant synchronization time window before the system eventually reaches the stationary state. A separate discussion is needed in the presence of noiseless and decoherence-free subspaces (this case is considered in Refs.\cite{manzanoSciRep,Cabotnetw}) and in Sect. 4.2.).

\subsubsection{Spin systems}
 
When considering discrete systems, like spin networks, or bosonic and fermionic 
systems in the presence of just one excitation,
the dynamics of the reduced system can be conveniently studied introducing the 
Liouville representation of the density matrix, which can be mapped into a 
vector belonging to a  Hilbert-Schmidt space
\begin{equation}
\rho = \sum_{i,j} \rho_{ij} |i\rangle \langle j| \rightarrow \vert\rho \rrangle
= \sum_{i,j} \rho_{ij}  \Ket{ij } ,
\end{equation}
where $ |ij \rrangle= |i\rangle \otimes |j \rangle$. In such space, the inner product is defined as
 $\llangle \tau \vert\rho \rrangle =
\tr(\tau^\dagger \rho)$. Furthermore, for any operator  $O$
we also have
\begin{equation}
\Ket{ O \rho }  =  O \otimes \mathbb{I} \Ket{\rho }, \quad
\Ket{\rho O }  =  \mathbb{I} \otimes O^\transp \Ket{\rho}
\end{equation}
where $O^\transp$ is the transpose of $O$ and $\mathbb{I}$ is the
 identity matrix. This formalism allows one to write  the Liouville
representation of the master equation (\ref{MasterEquation}) as 
\begin{equation}
|\dot \rho
\rrangle = \mathcal{L} |\rho \rrangle,
\end{equation}
where  the Liouvillian $ \mathcal{L}$ can be constructed as
\begin{eqnarray}\label{CostruzioneL}
& & \mathcal{L}  =  -i \left(H \otimes \mathbb{I} - \mathbb{I}
\otimes H^\transp \right)   + \sum_{\mu}
 \left[2 L_\mu \otimes  ( L_\mu^\dag)^\transp - L_\mu^\dag L_\mu  \otimes \mathbb{I} -\mathbb{I}  \otimes  (L_\mu^\dag L_\mu
)^\transp \right]. \nonumber
\end{eqnarray}
The Liouville representation of the master equation can be interpreted as a Schr\"{o}dinger
equation whose dynamics is governed by a non-Hermitian generator $ \mathcal{L}^\dagger \neq
\mathcal{L}$.

Using the Liouville formalism has the great advantage that a direct inspection of its spectrum is in general possible and then it is possible to identify the regions of parameters where synchronization is expected to come out.   Given an arbitrary initial state $\Ket{\rho_0}$,
the evolved density matrix can be written  as
\begin{equation}\label{rhot}
  \Ket{\rho_t}=\sum_{i} p_{0\, i}\, \Ket{ \tau_i} \,
\mathrm{e}^{\lambda_i t},\end{equation}
where $\Ket{\tau_i}$ are the right eigenvectors of $\mathcal{L}$, $\llangle
\bar{\tau}_i|$ are its left eigenvectors, and $ p_{0\,
i}=\frac{\llangle \bar{\tau}_i \Ket{ \rho_0}}{\llangle
\bar{\tau}_i\Ket{\tau_i}}$. Under these premises, a sufficient condition is the following: transient  synchronization emerges if  there exists a Liouvillian eigenvalue $\lambda_{\bar{i}}$ such that,  for any $i\neq \bar{i}$, $|{\rm Re}[\lambda_{\bar{i}}]|\ll |{\rm Re}[\lambda_i]|$
\cite{GLG2,Bellomo}. We remind that ${\rm Re}[\lambda_i]\leq 0$ $\forall i$, as they correspond to the decay rates of the Liouvillian eigenvectors. 
The condition is actually not necessary, as in a multipartite setting it can also happen that more than one mode share almost the same decay rate. In this case, the discriminant for the observation of synchronization would be the frequencies of such modes: if identical, a kind of macroscopic (almost) monochromatic oscillation would be observed, leading to collective synchronization; if different, synchronization would be hindered \cite{Cabot_atoms}.  

 The  Liouville formalism is convenient in the case of discrete variables, where the vector $| \rho
\rrangle$ is finite,  and will be adopted when dealing with coupled spins. In the case of  continuous variables, the problem would be 
much harder to be treated, a remarkable exception being represented by the evolution of Gaussian states, when it is sufficient to consider the covariance matrix (of finite dimensions) instead of the full density matrix.

\subsubsection{Harmonic networks}

A microscopic approach to establish the transient SS conditions of time scale separation has been reported in \cite{manzanoSciRep,Cabotnetw}
and follows from the analysis of the system-bath interaction Hamiltonian $H_I$,  which determines
the Lindblad master equations in the weak coupling limit for separate, common, and local baths. 
In fact,
looking at the microscopic model leading to the master equation
(\ref{MasterEquation}) and at the normal modes diagonalizing the system, one can infer the presence of a slowest mode, as this will be  weakly coupled to the environment.

The Hamiltonian of the system in the case of a harmonic network, $H_S 
 = \frac{1}{2} \left( \bf{p}^\transp {\bf{p}} + {\bf{x}}^\transp \mathcal{H}  {\bf{x}} \right)$, 
 can be diagonalized in the basis of its eigenmodes $\bf{X}=\mathcal{F}^\transp\bf{x}$ and $\bf{P}=\mathcal{F}^\transp\bf{p}$.
 In a microscopic description with independent oscillators modeling the
environment, the system-bath interaction Hamiltonian for separate baths (SB) takes the form
\begin{equation}\label{HISB}
H_I^{SB} = -  \sum_{m=1}^{N}  x_m B_m \text{  , with }
B_m=\sum_{k=1}^{\infty}s_k^{(m)} Q_k^{(m)},
\end{equation}
being
$Q_{k}^{(m)}$ the position operators for each
environment oscillator $k$ of the bath $B_m$,  and $s_k^{(m)}$ the coupling strength of $Q_{k}^{(m)}$ with the system oscillator $x_m$.
Supposing that there is a dominating channel of dissipation in one node, i.e. the coupling between oscillator $x_M$ and bath $B_M$ is significantly larger than the rest, then 
\begin{equation}\label{HILB}
  H_I^{LB} =- x_M B_M=  -  \sum_m {\kappa_m} X_m B_M \text{  , with }
  {\kappa_m} = \mathcal{F}_{M m}.  
\end{equation}
Therefore  this local bath (LB) configuration allows for imbalanced losses of the normal modes $X_m$.
On the other hand,  if we consider equivalent SBs, for which the system-bath couplings are the same for all system  oscillators, $s_k^{(m)}=s_k$ $\forall k$ and for each $m$, then all eigenmodes are coupled with the same strength to an independent bath, no matter the topology and characteristics of the system network, and no time scale separation would occur, provided the spectral density is flat enough. As these examples  (of local dissipation and equivalent SBs) show, the presence of multiple dissipative time scales and of transient SS will depend crucially on the coupling strengths of the oscillators to their baths. 

An interesting case discussed in detail in  \cite{Cabotnetw} is when  a common bath (CB)
 is seen by all oscillators in the network, with an interaction Hamiltonian
 \begin{equation}\label{HICB}
  H_I^{CB} = -  \sum_{m=1}^{N}   x_m B \text{  , with }
B=\sum_{k=1}^{\infty}c_k Q_{k},
\end{equation}
that involves only the center
of mass of the network and a single bath. Here we denote the coupling strength of the oscillators with the bath modes as $c_k$. 
In fact, the master equation for  CB can also be obtained from   SB one assuming perfect correlations between the different environments. It is illustrative to rewrite (\ref{HICB}) in the eigenmode basis
\begin{equation}\label{HICB_nm}
  H_I^{CB} =  -  \sum_m \kappa_m X_m B \text{  , with }\kappa_m = \sum_n  \mathcal{F}_{n m}.
\end{equation} 
In both Eqs. (\ref{HILB}) and (\ref{HICB_nm}), the \textit{effective couplings} $\kappa_m$ are different and determined by
characteristics of the network such as topology, coupling
strengths, and frequencies, as encoded in the diagonalization matrix $\mathcal{F}$. These effective couplings characterize the dissipation rate of the eigenmodes. Hence,  we can readily assess the possible emergence of SS analyzing the relative magnitudes of these couplings: significant time scale separation is anticipated by one $\kappa_m$ being significantly smaller than the rest \cite{manzanoSciRep,Cabotnetw}. It can also happen that one eigenmode is indeed perpendicular to the center of mass vector, which leads to $\kappa_m=0$, and thus to this eigenmode being effectively uncoupled from the bath and immune to dissipation and decoherence. The physical consequences of this are analyzed in detail in Refs. \cite{manzanoPRA,manzanoSciRep,Cabotnetw} and will be further unfolded in Subsec. \ref{sec42}.

\section{Quantifying transient synchronization}\label{seciii}

The general problem of quantifying synchronization in the quantum regime was tackled in Ref. \cite{ZambriniRev}, where a discussion about local and global indicators (the latter ones encompassing  various kinds of  correlations) was presented. 
Local indicators must be able to  detect linear dependence between time-dependent variables. The most commonly used of this kind is the Pearson's  correlation coefficient
first considered for quantum synchronization in Ref. \cite{SyncHO}.  Given two time-dependent variables $A_1$ and $A_2$, the
Pearson's  parameter ${\cal C}_{A_1,A_2}(t|\Delta t)$ can be calculated over a
sliding window of length $ \Delta t$:
\begin{equation}\label{eq:pears}
 {\cal C}_{A_1,A_2}(t|\Delta t)=\frac{\int_{t}^{t+\Delta t}(A_1-\bar{A_1})(A_2-\bar{A_2})dt}
 {\sqrt{ \int_{t}^{t+\Delta t}(A_1-\bar{A_1})^2 dt  { \int_{t}^{t+\Delta t}(A_2-\bar{A_2})^2 dt}}}
\end{equation} 
with
\begin{equation}
 \bar{A_1}=\frac{1}{\Delta t}\int_{t}^{t+\Delta t}A_1dt
\end{equation} 
and $A_i$ are expectation values of quantum operators. 

As for collective indicators, one could consider two- or many-body quantum correlations to assess the presence of SS. In Ref. \cite{ZambriniRev} it was questioned whether,  for instance, spin-spin correlations of the form ${\rm Re}[\langle \sigma_i^-(t) \sigma_j^+(t)\rangle]$ are a faithful indicator. Based on the spectral analysis given in the previous section, one can deduce that the presence of such correlations  actually represents a sufficient condition for synchronization, as it would witness the presence of a slow decaying mode.  
On the other hand, invoking the quantum regression theorem \cite{Carmichael}, one can see that the same spectral decomposition found in local observables can also be found in two-time, steady-state two-body correlation functions.
As an alternative, in Ref. \cite{ameri}, it was proposed to make use of the mutual information in the steady state to trace back the transient synchronization. Actually, as shown in  \cite{ZambriniRev}, it is not always the case, as there are scenarios where mutual information is totally unable to capture the transition from the presence to the absence of synchronization. 
In the following the emergence of SS will be discussed through the Pearson indicator (\ref{eq:pears}) , being this the most restrictive characterization for SS.

\section{Examples} \label{seciv}

\subsection{A pair of harmonic oscillators}
The first model where transient synchronization was studied is represented by coupled harmonic oscillators in the presence of either separate environments or a common one \cite{SyncHO}. The simplest of such systems displaying SS in the relaxation dynamics is represented by
two elements with equal mass ($m=1$) and different frequency:
\begin{equation}
H_S=\frac{p_1^2}{2}+\frac{p_2^2}{2} 
+\frac{\omega_1^2}{2}x_1^2+\frac{\omega_2^2}{2}x_2^2 +\lambda x_1x_2.
\end{equation}
Dissipation is modeled through a microscopic model given in Eq. (\ref{HISB}) for SB and   Eq. (\ref{HICB}) for CB.
From these Hamiltonians we can derive the standard master equations describing the dissipative dynamics of the system. Hence, assuming identical baths, and under the standard Born-Markov, secular approximation, the following master equation is found for both SB and CB in the normal mode basis and in the global approach  \cite{SyncHO,manzanoPRA,manzanoSciRep}:
\begin{eqnarray}
\dot{\rho}(t)&=&-\frac{i}{2}[P_+^2+P_-^2+\Omega_+^2 X_+^2+\Omega_-^2X_-^2,\rho(t)]
-\frac{1}{4}\sum_{n=\pm}i\Gamma_n\big([X_n,\{P_n,\rho(t)\}] \nonumber\\&-&[P_n,\{X_n,\rho(t)\}]\big)
+D_n\big([X_n,[X_n,\rho(t)]] -\frac{1}{\Omega_n^2}[P_n,[P_n,\rho(t)]]\big),
\end{eqnarray}
%
where the expressions of the normal modes in terms of the coupled oscillators  can be found in \cite{SyncHO}. Notice that the SB and CB cases differ in the parameters $\Gamma_\pm$ and $D_\pm$ of the master equation. While in the SB case we find the decay rates, $\Gamma_\pm$, to be identical for both normal modes, in the CB case they are weighted by the normal mode shape $\kappa_m=\sum_n \mathcal{F}_{nm}$. In particular, for SB we have
\begin{equation}
\Gamma_\pm=\gamma, \quad D_\pm=\gamma \Omega_\pm \coth \bigg(\frac{\Omega_\pm}{2T}\bigg),    
\end{equation}
while for CB
\begin{equation}
\Gamma_\pm=\gamma\kappa_\pm^2, \quad D_\pm=\kappa_\pm^2\gamma \Omega_\pm \coth \bigg(\frac{\Omega_\pm}{2T}\bigg).   
\end{equation}
where we have introduced the phenomenological parameter $\gamma$ for the strength of dissipation. Notice, however, that an expression for $\gamma$ in terms of the microscopic parameters of the system-bath model can be obtained \cite{bp}. This difference in the decay rates ultimately leads to the emergence of phenomena as transient synchronization or decoherence free evolution, as we will see in the following. 

\begin{figure}[H]
 \centering
 \includegraphics[width=0.95\columnwidth]{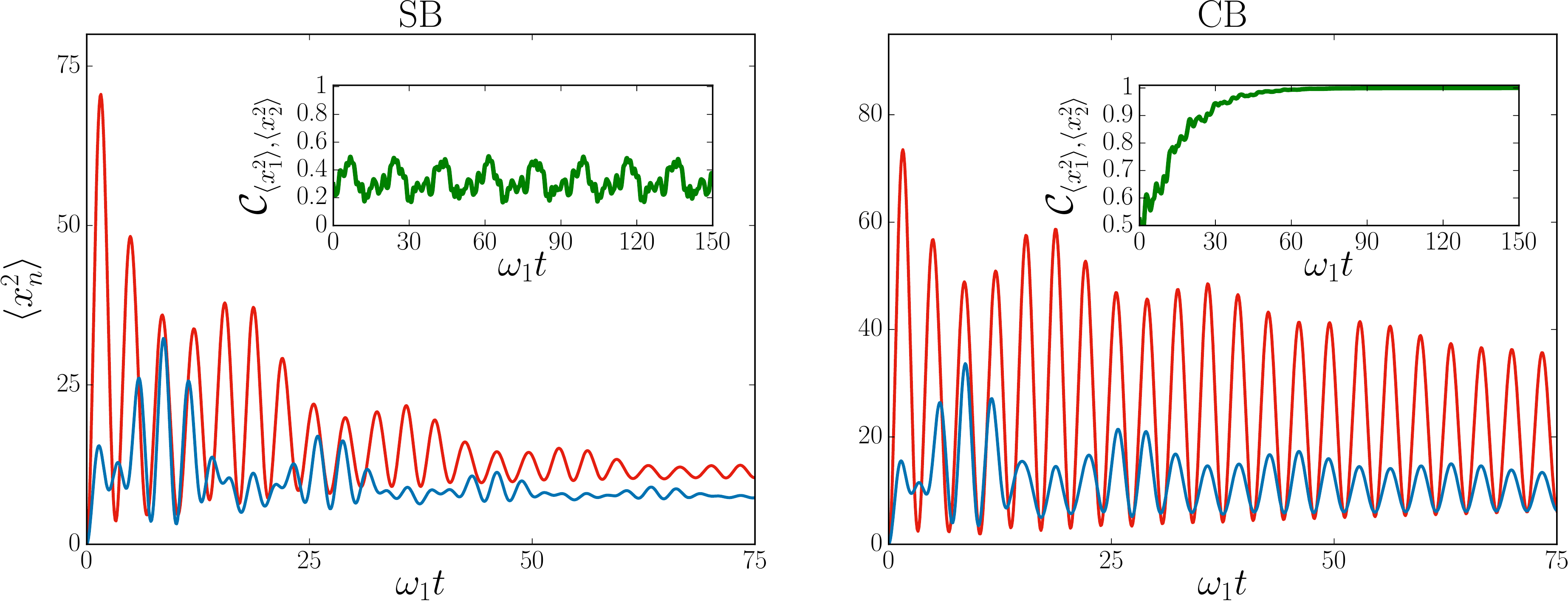}
 \caption{In red $\langle x_1^2(t)\rangle$, in blue $\langle x_2^2(t)\rangle$ for SB (left panel) and CB (right panel), with $\omega_2/\omega_1=1.2$, $\lambda/\omega_1^2=1.3$, $\gamma/\omega_1=0.05$ and $T/\omega_1=10$.  The initial condition is a separable vacuum state with $r_1=2.5$ and $r_2=1.8$. In the insets we plot the synchronization measure: $\mathcal{C}_{\langle x_1^2\rangle,\langle x_2^2\rangle}(\omega_1 t|\omega_1\Delta t=20)$.}
 \label{sync_emergence}
\end{figure}

\begin{figure}[H]
 \centering
 \includegraphics[width=1.0\columnwidth]{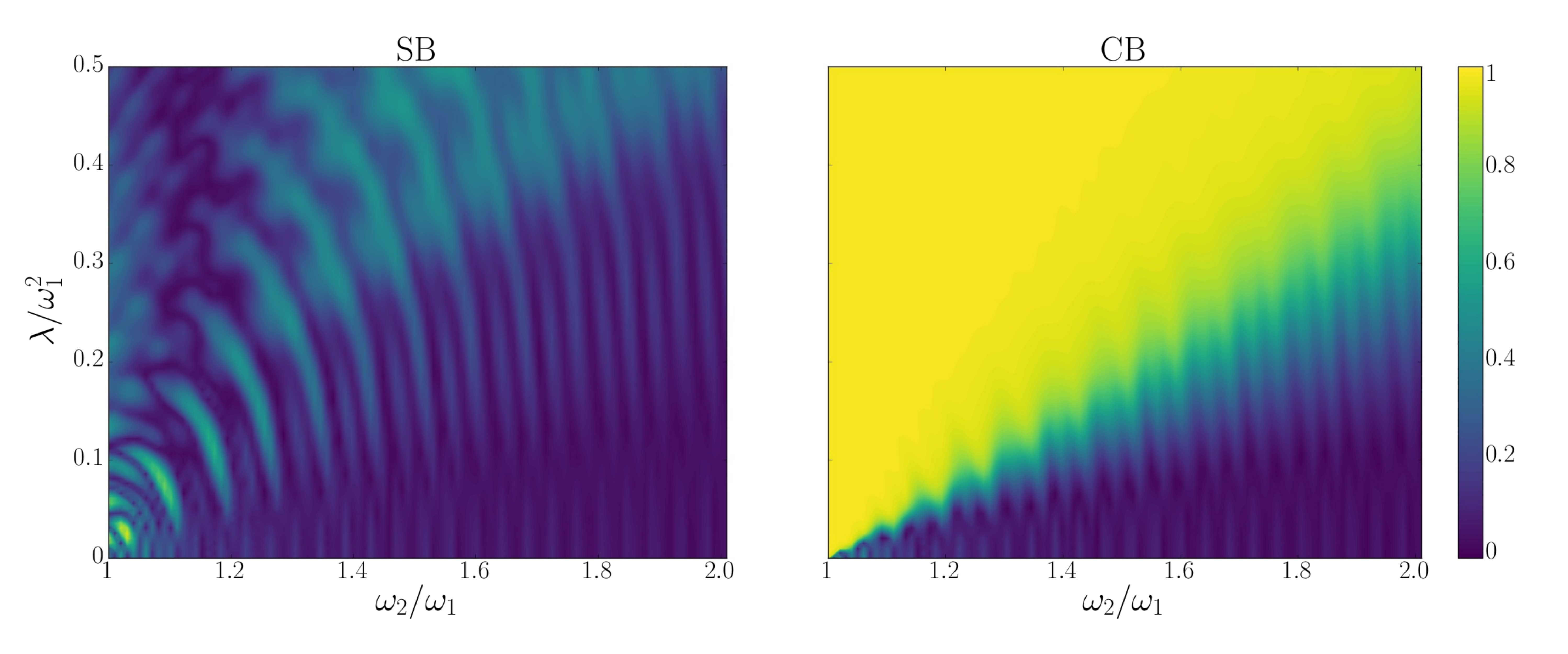}
 \caption{In color  $|\mathcal{C}_{\langle x_1^2\rangle,\langle x_2^2\rangle}|$ at $\omega_1 t=70$ and for a time window $\omega_1 \Delta t=20$. Here we vary the coupling strength $\lambda/\omega_1^2$ and frequency $\omega_2/\omega_1$. The initial condition is a separable vacuum state with $r_1=2$ and $r_2=1$. We fix $T/\omega_1=10$ and $\gamma/\omega_1=0.05$.}
 \label{tongue}
\end{figure}

The linearity of the dynamics, coming from the fact that the Hamiltonian is quadratic, implies  that if the initial state  is Gaussian,
it will remain so at all times. Then, in this case, the dynamics  can be fully  described only looking at the first and second moments of the system, whose dynamical equations are derived from the above master equation and can be found in \cite{manzanoSciRep}. For instance, synchronization can  be found in the time evolution of second moments of a vacuum squeezed state
\begin{equation}
\langle x_n^2(0)\rangle=\frac{e^{-2r_n}}{2\omega_n},\quad \langle p_n^2(0)\rangle=\frac{\omega_n e^{2r_n}}{2},\quad n=1,2,   
\end{equation}
the rest of first and second moments being initially zero.  As observed in Ref.  \cite{SyncHO}, if the baths are separate and identical, synchronization never emerges. In Fig. \ref{sync_emergence}, we compare CB and SB  and show  how synchronization emerges in one case and not in the other. Notice that besides $\langle x_1^2(t)\rangle$, $\langle x_2^2(t)\rangle$, synchronization is observed in the other second moments too. In Fig. \ref{tongue} the Pearson indicator is drawn for different detunings and coupling strength. We observe how synchronization never emerges in the SB case, despite increasing the coupling strength for a given detuning. On the contrary, in the CB case we observe the typical Arnold tongue behavior: as coupling strength increases synchronization emerges for larger detunings.

\subsection{From transient to stationary synchronization} \label{sec42}

Beyond transient SS, when considering more than two detunded oscillators, in the CB case it can happen that one or more normal modes are effectively uncoupled from the bath, i.e. $\kappa_m=0$. These states  are dark with respect to the noise and evolve freely. Then, depending on the initial conditions, the system does not
fully thermilize and the asymptotic state itself can exhibit a synchronous dynamics and asymptotic correlations. This stationary synchronization
due to the presence of decoherence free subspaces was shown in detail in small systems of coupled harmonic oscillators in \cite{manzanoPRA,manzanoSciRep}, while in \cite{Cabotnetw} the focus was set on  larger networks with complex topologies such as the Erd\H{o}s-R\'{e}nyi or Small-World networks. One of the main results found in \cite{Cabotnetw} was that parameter and topological uniformity are crucial for the presence of these {\it noisless modes}. Indeed, it was shown that in extended networks the parameter regions with larger degree variance coincided with the regions with less probability of having noisless modes. In contrast, in this same work, it was found that the conditions for transient synchronization to emerge were largely met, despite parameter and topological disorder.

\begin{figure}[H]
 \centering
 \includegraphics[width=0.95\columnwidth]{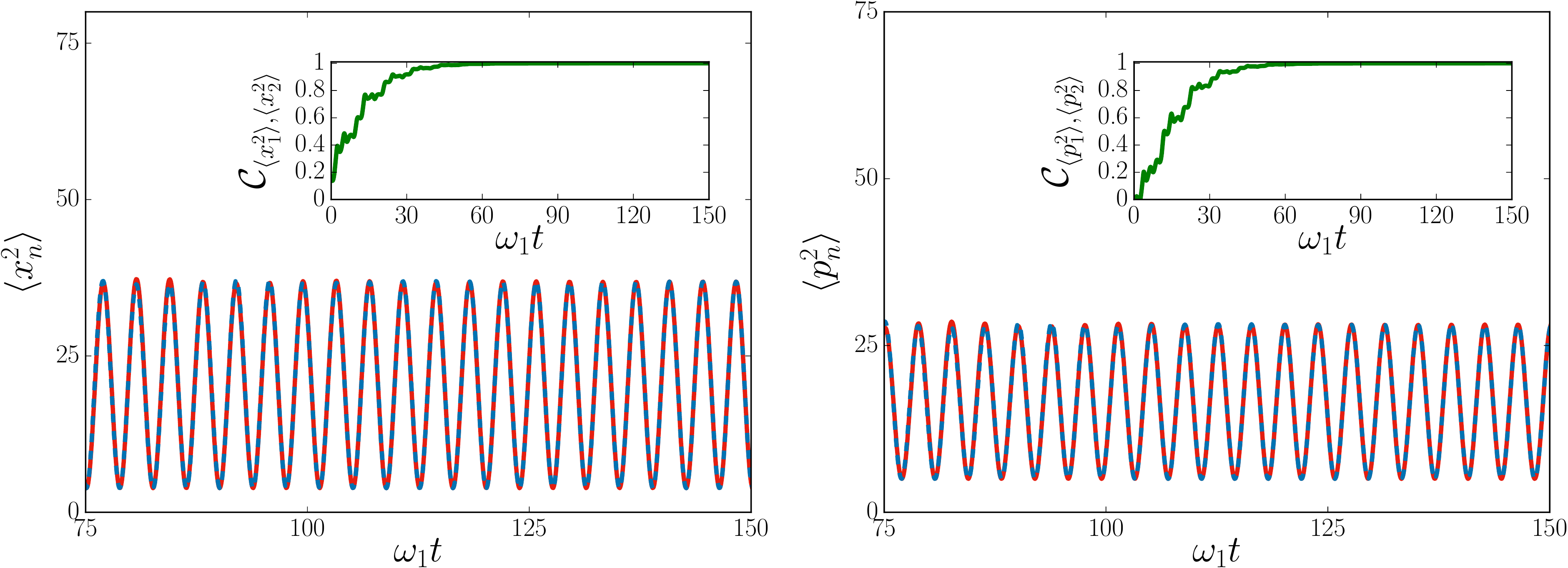}
 \caption{Example of stationary synchronous state. In red $\langle x_1^2(t)\rangle$ (left panel) $\langle p_1^2(t)\rangle$ (right panel), in dashed blue lines $\langle x_2^2(t)\rangle$ (left panel) $\langle p_2^2(t)\rangle$ (right panel), in the CB case with $\omega_2/\omega_1=1.0$, $\lambda/\omega_1^2=0.3$, $\gamma/\omega_1=0.05$ and $T/\omega_1=10$.  The initial condition is a separable vacuum state with $r_1=2.5$ and $r_2=1.8$. In the insets we plot the synchronization measure: $\mathcal{C}_{\langle x_1^2\rangle,\langle x_2^2\rangle}(\omega_1 t|\omega_1\Delta t=20)$ (left panel) and $\mathcal{C}_{\langle p_1^2\rangle,\langle p_2^2\rangle}(\omega_1 t|\omega_1\Delta t=20)$ (right panel)  .}
 \label{fig3}
\end{figure}

Stationary synchronization due to the presence of decoherence free subspaces arises also in the (simplest) case of two
coupled identical oscillators ( $\omega_2=\omega_1$). In the CB case, $\kappa_-=0$ implying that the evolution of $X_-$ is only ruled by the Hamiltonian. Then, for any initial condition overlapping with $X_-$, the dynamics of the system will be initially a mixture of modes, but eventually, only this non-decaying mode will survive, resulting in asymptotic synchronous oscillations as shown in Fig. \ref{fig3}.

\subsection{Spin pairs: local versus global environment}\label{spinpairs}

As discussed so far, SS can be induced by the presence of a common environment which selects the slowest decaying mode. This mechanism for SS is perhaps the one that shares more features with synchronization in the classical domain. In fact, as we have already seen, synchronization is favoured by a small detuning among the different units, while it is hindered in the case where the local energies are different enough. Actually, as discussed in Refs. \cite{GLG2,Cabot_atoms}, the quantum realm offers yet another mechanism where the emergence of transient synchronization in the presence of local environments is made possible by a finite energy detuning among local components, while it is suppressed in the small-detuning limit. In this scenario, similar to what observed in the case of nonlinear quantum harmonic oscillators \cite{blockade}, the enabling factor is represented by the imbalance among the losses of any local bath. 

For the sake of clarity, let us consider two non-interacting qubits in the presence of a local  bath \cite{GLG2}: the system is described by
 \begin{eqnarray}
 H_S=\frac{\omega_1}{2}\sigma_1^z+\frac{\omega_2}{2}\sigma_2^z+\lambda \sigma_1^x 
\sigma_2^x, \label{hs}
 \end{eqnarray}
while the interaction with the bath is
\begin{equation}
H_I^{ LB}=  \sum_k g_k (a_k^\dag+ a_k )\sigma_1^x.
\end{equation} 
Indeed, the fact that the interaction with the environment is local does not necessarily imply that  the master equation is local itself. In fact, following the standard approach \cite{bp}, we first need to find the normal modes of $H_S$ and then write the dissipator using the jump operators among such modes. Hamiltonian (\ref{hs}) can be diagonalized through a Jordan-Wigner transformation \cite{lsm}, leading to 
 \begin{equation}
 H_S= E_1 (\eta_1^\dag \eta_1-1/2)+E_2 (\eta_2^\dag \eta_2-1/2),
 \end{equation}
 with $\eta_i$ fermionic annihilation operators defined through
 \begin{eqnarray}
     \sigma_1^+&=&\cos{\theta_+}(\cos \theta_- \eta_1^\dag+\sin\theta_-  \eta_2^\dag)+\sin{\theta_+}(\cos \theta_- \eta_2 -\sin\theta_-  \eta_1),\nonumber\\
     \sigma_2^+&=&(1-2  \eta_1^\dag \eta_1)[\cos{\theta_+}(\cos \theta_- \eta_2^\dag -\sin\theta_-  \eta_1^\dag)-\sin{\theta_+}(\cos \theta_- \eta_1+\sin\theta_-  \eta_2)],\nonumber\\
 \end{eqnarray}
where
\begin{equation}
 \theta_\pm=\frac{1}{2}\arcsin\frac{2\lambda}{\sqrt{4\lambda^2+(\omega_1\pm\omega_2)^2}}.
\end{equation}
As the interaction term can be also written as 
\begin{equation}
 \sigma_1^x =\cos (\theta_++\theta_-)(\eta_1^\dag+ \eta_1)+\sin(\theta_++\theta_-)(\eta_2^\dag+\eta_2),\label{sq}
 \end{equation}
 the master equation (for simplicity we assume temperature $T=0$) will take the  form 
 \begin{equation}
\dot{\rho}=-i[H_S,\rho]+\sum_{i=1}^2\tilde\gamma_i (\eta_i \;\rho \;\eta_i ^\dag-\{\rho,\eta_i ^\dag\eta_i\})+\sum_{i,j=1}^2\tilde\gamma_{ij} (\eta_i \;\rho \;\eta_j ^\dag-\{\rho,\eta_j ^\dag\eta_i\}),\label{diss}
\end{equation} 
with $\tilde\gamma_1=\cos^2 (\theta_++\theta_-)J( E_1)$ and  $\tilde\gamma_2=\sin^2 (\theta_++\theta_-)J( E_2)$, where $J(\omega)=\sum_{k}|g_k|^2\delta(\omega-\omega_k)$ is the spectral density of the bath which characterizes microscopically the system-bath coupling \cite{GLG2,bp}. The final term on the right-hand side of Eq.  (\ref{diss}) takes into account contributions that are not associated to resonant jump but may not be negligible with respect with the full secular terms. The weight of such terms is expected to vanish for large values of $\lambda$. We remark here that such a partial secular approximation still guarantees that the master equation has a Lindblad form  \cite{marco,facer,giovannetti}.
It is clear from the structure of the master equation that synchronization can only take place if the dissipation rates of the two normal modes $\eta_1$ and $\eta_2$ are significantly different between each other and the modes themselves have a finite superposition over the two local spins. 

Despite the fact that the coupling with the bath was local, the master equation (\ref{diss}) has a clear non-local form, because of the non-local character of the fermionic mode operators.  In general the jump operators are non-local  for any finite value of $\lambda$, while they can converge to local ones only in the limit $\lambda\to 0$. On the other hand, in the limit where the baths are separate and the local picture can give a good approximation of the true dynamics, SS is not expected to emerge. 

\begin{figure}[H]
 \centering
 \includegraphics[width=0.95\columnwidth]{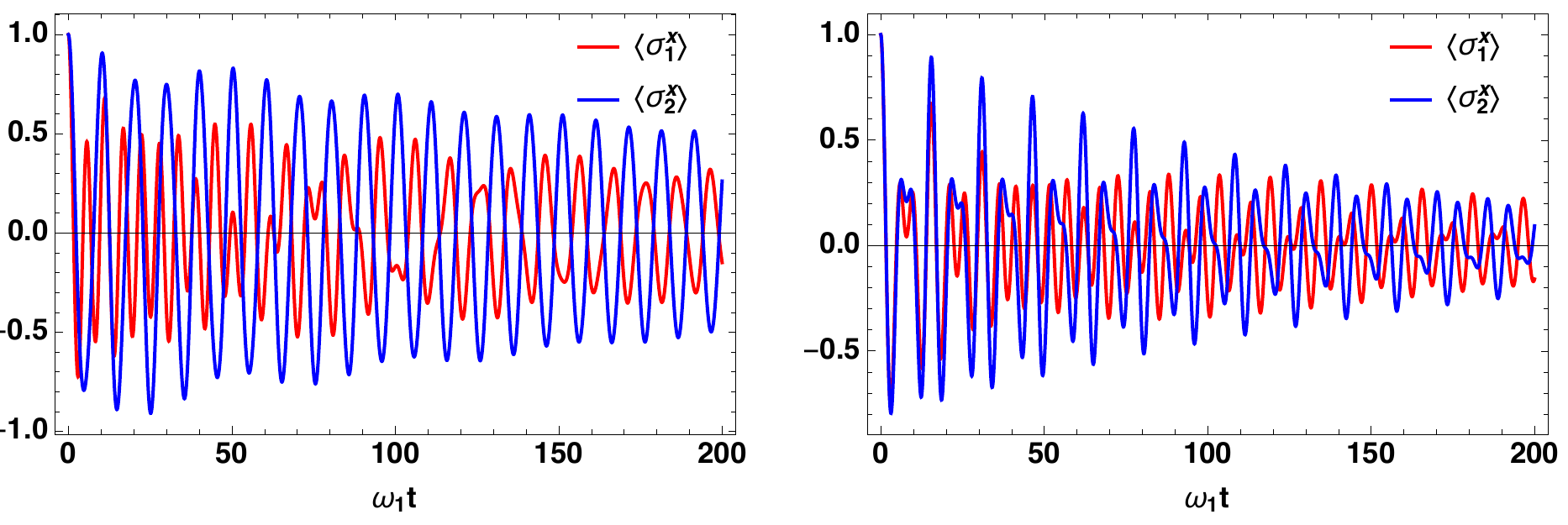}
 \caption{In red $\langle \sigma^x_1(t)\rangle$, in blue $ \langle\sigma^x_2(t)\rangle$ for two different choices of parameters. Left: $\omega_2=0.7 \;\omega_1$; right: $\omega_2=0.99\; \omega_1$.}
 \label{figsec43}
\end{figure}

In Fig. \ref{figsec43}, we show the role played by the detuning for such a system. Taking an Ohmic environment at zero temeprature with system-bath coupling $\gamma_0=5\cdot10^{-3}\omega_1$ and a spin-spin coupling $\lambda=0.2 \omega_1$, we plot the coherences of the two qubits for large detuning ($\omega_2=0.7 \omega_1$, left panel) and for small detuning  ($\omega_2=0.99 \omega_1$, right panel).  Synchronization is forbidden in the region where the two frequencies are very close to each other, while it emerges in the case of larger detuning. The scenario proposed here is extreme, as only one of the two qubits has direct interaction with an environment. Synchronization is also possible in the case where both qubits dissipate locally, but with different local decay rates \cite{Bellomo}.

 \subsection{The role of decoherence in transient synchronization}

The previous discussion indicates that the separation of time scales induces synchronous dynamics. In all the examples, we have assumed a dissipative mechanism bringing the density matrix to its equilibrium steady-state. Actually, this is not the only way an extended environment can affect the evolution of a quantum state. Another important mechanism, commonly known as phase damping channel \cite{nielsen}, generates pure dephasing among the eigenstates of the system Hamiltonian without affecting their initial populations. It takes place whenever $[H_S,H_I]=0$, where $H_S$ is the system Hamiltonian and $H_I$ describes the system-bath interaction.

In general, dissipation and dephasing can coexist, and $H_I$ can be written as the sum of two terms, each of them accounting for the two distinct kinds of noise. The interplay between these two mechanisms was discussed in Ref. \cite{GLG1} where it was shown that actually pure dephasing does not  favor the emergence of SS. A deeper analytical study of the phenomenon can be  performed looking at the properties of the  Liouvillian spectrum. Let us consider the case of two non-interacting spins, detailed in Ref. \cite{Bellomo}:
\begin{equation}
 H_S=\frac{\omega_1}{2}\sigma_1^z+\frac{\omega_2}{2}\sigma_2^z
\end{equation}
Assuming a dissipative bath with decay rates $\gamma_{ij}$ and a dephasing bath $\gamma_{ij}^z$, the master equation is
\begin{equation}
    \dot{\rho}=-i[\tilde H_S,\rho]+2\sum_{i,j=1}^2 \gamma_{ij}\left(\sigma_i^-\rho \sigma_j^+ -\{\rho,\sigma_j^+\sigma_i^-\right\})+2\sum_{i,j=1}^2 \gamma_{ij}^z\left(\sigma_i^z\rho \sigma_j^z -\{\rho,\sigma_j^z\sigma_i^z\}\right)
\end{equation}
where $\tilde H_S$ is the system Hamiltonian renormalized by a Lamb shift term that introduces an effective coupling between the two spins, $H_{LS}\simeq s (\sigma_1^+\sigma_2^-+h.c.)$ \cite{Bellomo}. Here the bath-induced coherent interaction is treated phenomenologically  and it is parametrized by the coupling rate $s$. Notice that an expression for $s$ in terms of the parameters of the microscopic system-bath model can be obtained \cite{bp}. In the case of  common baths, we have $\gamma_{ij}= \gamma$ and $\gamma^z_{ij}= \gamma^z$  for any $i,j$. Due to the symmetry of the problem, the Liouvillian superoperator is the direct sum of five independent block, and each of them can be studied separately. The spin coherences $\sigma_x^i$ (or, equivalently $\sigma_y^i$) used to determine synchronization belong to the subspaces  whose evolution is determined by 
\begin{equation}\label{lc}
{\cal L}_c=
    \left(\begin{array}{cccc}
        -3 \gamma-2 i \omega_2  &  -\gamma+i s & 0&0\\
        -\gamma+i s & -3 \gamma-2 i \omega_1 & 0&0\\
        2 \gamma &2 \gamma & - \gamma-2 i \omega_1& -\gamma-i s\\
         2 \gamma &2 \gamma & -\gamma-i s & -\gamma-2 i \omega_2
    \end{array}\right)-4 \gamma_z \mathbb{I}_4,
\end{equation}
where  $ \mathbb{I}_4$ is the identity operator in the $4\times 4$ space, together with its conjugate ${\cal L}_c^*$. Equation (\ref{lc}) immediately tells us what is the effect of pure dephasing: it introduces a shift of all the eigenvalues, which means that the decoherence dynamics is homogeneously accelerated. Thus, the difference between (the two slower) pairs of eigenvalues is not affected while their ratio is diminished. Then,  the amplitude of the slowest eigenmode is damped by a factor $e^{-4 \gamma_z t}$. This also implies that  pure dephasing itself would never be able to induce any synchronization. We mention here the fact that, in the presence of direct coupling between the two spins, a spin-bath interaction proportional to $\sigma_z$ (it can be either local or global), would not represent a pure dephasing channel, and could then contribute to the emergence of SS.

Another consequence about the presence of a phase damping channel concerns the fact that it breaks the complete correspondence between SS and the collective emission of radiation observed in Ref. \cite{Bellomo}. Indeed, superradiance \cite{haroche} can be calculated taking the intensity of the emitted radiation according to
\begin{equation}
    I(t)=\sum_{i,j}\Gamma_{i,j}\langle \sigma_i^+(t)\sigma^-_j(t)\rangle,
\end{equation}
 where $\Gamma_{i,j}$ take into account any decoherence process. However, at least in the case of a common dephasing bath, $I(t)$ is left totally unchanged by the presence of such bath. As discussed in Ref. \cite{Bellomo}, in the  absence of a pure dephasing channel, the time scales of spontaneous synchronization and of the emergence of subradiant emission are equal, while this perfect matching is broken by the presence of dephasing.

\section{Conclusions }\label{secv}
 Synchronization in the quantum regime can emerge as a long-time dynamical regime before the system reaches the final steady state. In such a  transient, the expectation values
of the system operators  present synchronous oscillations. While synchronization is measured taking temporal correlations of classical trajectories of local observables, the true quantum nature of this phenomenon can be traced back to the fact that we have considered quantities that would not be present in the classical case, as they directly derive from the existence of quantum coherence.  We have considered both finite systems (qubits) and harmonic oscillators, reviewing the general conditions that must be fulfilled in order to observe the emergence of SS.

Transient synchronization  presents various interesting aspects that have been reviewed here.  Depending on the geometry of the system-bath interaction, synchronization can be either enhanced or suppressed by the amount of detuning among the local units. This is quite uncommon in classical synchronization scenarios, where, normally, large values of the detuning hamper the possibility of observing synchronization. 
We have also discussed the role played by the presence of either local or collective losses and the possibility of having persistent synchronous steady-state oscillations due to the presence of noiseless modes in the dynamics. Finally, we have discussed the effect of pure dephasing,  discussing its effect on transient synchronization, and the connection of the latter to super- and sub-radiance. 

 \subsection*{Acknowledgements}

The authors acknowledge support from MINECO/AEI/FEDER through the project EPheQuCS FIS2016-78010-P, the Mar\'ia de Maeztu Program
for Units of Excellence  (MDM-2017-0711), and
funding from CAIB PhD and postdoctoral programs.

\end{document}